# Enhanced spin hall effect of reflected light due to Optical Tamm states with Dirac semimetal at the terahertz range

Keqiang Yin[a], Luzihao Li[a,1], Qijun Ma[a], Jie Jiang[b], and Leyong Jiang[a,*]

[a]School of Physics and Electronics, Hunan Normal University, Changsha 410081, China;

[b]Hunan Key Laboratory of Super Microstructure and Ultrafast Process, School of Physics and Electronics, Central South University, Changsha 410083, China；

[*] Corresponding author.
E-mail address: jiangly28@hunnu.edu.cn
[1] K. Yin and L. Li contributed equally to this work.

**Abstract:** The enhanced photonic spin hall effect (PSHE) plays a positive role in the flexible manipulation of photons. Here, by combining Dirac semimetals with Bragg reflector constructed by one-dimensional photonic crystal, we theoretically design a simple multilayer structure to enhance and manipulate the PSHE. Through the detailed and optimal design of the conductivity characteristics of Dirac semimetals and the structural parameters of the whole model, we realize the excitation of Optical Tamm states (OTSs) in the multilayer structure, so that the PSHE can be enhanced and controllable. The theoretical results show that by optimizing the Fermi energy and the thickness of Dirac semimetal, the reflection coefficient ratio can be increased under p-polarization and s-polarization, thus creating conditions for enhanced PSHE. In addition, the effects of the incident angle and the parameters of the spacer layer on the PSHE are also clarified. We believe above results can provide a new paradigm for the construction of controllable spin devices.

# 1. Introduction

The photonic spin hall effect (PSHE) refers to the left and right-handed circular polarization components will produce a spin-dependent splitting shift in the center of gravity when the linearly polarized beam is reflected or refracted on the interface of the medium [1-4]. Some interesting applications of PSHE in optoelectronic devices have been proposed, such as barcode encryption [5], multichannel switch [6] and biosensor [7]. However, the spin splitting caused by the PSHE is very weak, and traditional optical equipment is not convenient for direct observation, which is not conducive to its application in micro and nano optoelectronic devices [8]. Therefore, researchers have proposed various ways to enhance the PSHE. For example, The research of Dong *et al.* indicated that we can enhance PSHE by controlling the optical properties of graphene through weak optical pumping at the terahertz range and the spin shift can reach 15 times the working wavelength [9]. Xiang *et al.* presented to enhance the optical transverse displacement by guided wave surface plasmon resonance and the transverse shift can reach 11.5 μm [10]. Kim *et al.* demonstrated that vertical hyperbolic materials have an advantage over horizontal hyperbolic materials in enhancing PSHE [11]. In terms of experimental observation, Qin *et al.* observed the spin splitting of the beam at the air-glass interface of several hundred nanometers by weak measurement method [12]. Luo *et al.* observed a large optical spin shift through weak measurement method when the beam reflected near Brewster angle [13]. In addition, various Dirac materials such as graphene [14, 15] and black phosphorus [16] have become hot spots of researching enhance and control the PSHE due to their excellent optical properties. Although many useful works have been done to enhance the PSHE in recent years but it is still worth exploring to seek the enhanced and easily controled PSHE.

For the past few years, with the increasing interest in Dirac materials people has been found that except two-dimensional Dirac materials, three-dimensional Dirac semimetals, "3D graphene" or bulk Dirac semimetals (BDS), such as $Na_3Bi$, $Cd_3As_2$, $AlCuFe$, have also attracted great attention [17-19]. Compared with graphene, BDS has more excellent optical properties, such as lower energy photon mobility and higher electron mobility [20-22], and BDS has lager volume, longer propagation length and more flexible design manufacture compared with graphene, so it will be widely used in the field of nano photons. Moreover, BDS also has dynamic and controllable conductivity [23], which provides a new way for us to design easily controled optoelectronic devices based for PSHE. In addition, OTSs as a lossless interface mode localized at the boundaries of two different periodic media [24], OTSs are more flexible than surface plasmons (SPPs). Because it can be both excited by p-polarization and s-polarization without no specific incident angle [25]. Graphene has been shown to excite OTSs at terahertz range [26, 27]. Therefore the enhancement and regulation have been proposed for PSHE by inspiring Tamm Plasmons (TPs) [28].

BDS as a new material with excellent optical properties like graphene, and the plasmonic waveguide excited at terahertz range has better confinement conditions and lower loss [29]. So it is feasible and attractive to excite OTSs based on BDS-distributed Bragg reflector structure to enhancing PSHE. In the work of this paper, we theoretically present a multilayer composite structure based on BDS-distributed Bragg reflector, and realize the enhancement and regulation of PSHE through exciting TPs. The enhancement of spin transverse shift is due to the TPs excitation at the BDS interface in the multilayer structure.In addition, we not only obtain the enhanced PSHE, but also the controllable conductivity of BDS provides a way for us to control the optical spin transverse shift based on this structure. We believe that the OTSs structure based on BDS will provide us more feasible options and potential applications in the field of optical spin devices.

## 2. Theoretical Model and Method

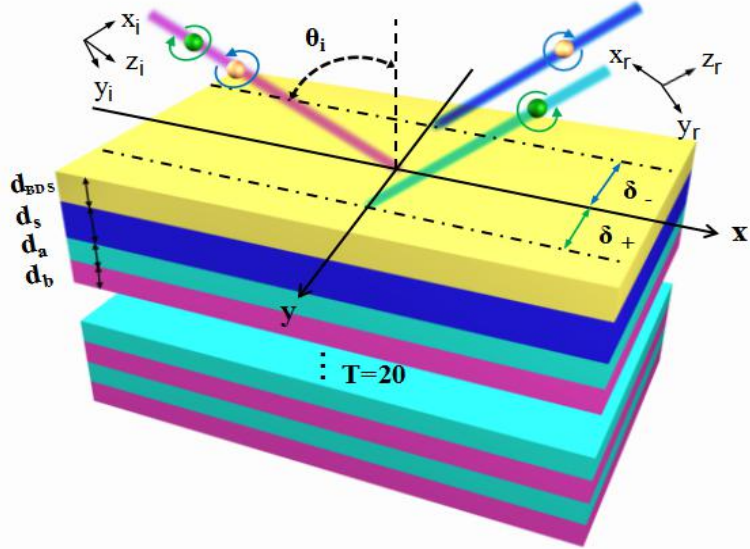

Fig. 1. Schematic diagram of the PSHE of the structure we proposed, where $d_{BDS}, d_s, d_a$ and $d_b$ represent the thickness of BDS, space layer, dielectric A and B respectively.

We consider a multilayer structure consisting of BDS, space layer and distributed Bragg reflector structure, as shown in Fig 1. The top layer of the multilayer structure is BDS, dielectric A and B form distributed Bragg reflector alternately with period T=20, dielectric A and B is poly-4-methyl pentene-1 (TPX), and SiO$_2$ respectively. $\lambda_c = 300$ μm is the central wavelength, the thickness of dielectric A and B is set as $d_{a,b} = \lambda_c / 4n_{4,5}$ [23], where $n_4 = 1.47$ is the refractive index of TPX, $n_5 = 1.90$ is the refractive index of SiO$_2$.

According to the random phase approximation theory, the dynamic conductivity of BDS can be expressed in Kubo formalism of the long-wavelength limit, similar to the two-dimensional Dirac material graphene. So the BDS conductivity is expressed as follows:[30] :

$$\mathrm{Re}\,\sigma(\Omega) = \frac{e^2}{\hbar} \frac{gk_F}{24\pi} \Omega G(\Omega/2), \tag{1}$$

$$\mathrm{Im}\,\sigma(\Omega) = \frac{e^2}{\hbar} \frac{gk_F}{24\pi^2} \left[ \frac{4}{\Omega} \left( 1 + \frac{\pi^2}{3} \left(\frac{T}{E_F}\right)^2 \right) + 8\Omega \int_0^{\varepsilon_c} \left( \frac{G(\varepsilon) - G(\Omega - 2)}{\Omega^2 - 4\varepsilon^2} \right) \varepsilon d\varepsilon \right], \tag{2}$$

where $e$ is the electron charge, $\hbar$ is the reduced Planck constant, $G(E) = n(-E) - n(E)$, $n(E)$ is the Fermi distribution function, $E_F$ is the Fermi energy, $v_F$ is the Fermi velocity, $\varepsilon_c = 3$, $\varepsilon = E/E_F$, $\Omega = \hbar\omega/E_F$ and $g$ is the degeneracy factor. When the conductivity of BDS is expressed, its permittivity can be expressed as follows:

$$\varepsilon = \varepsilon_b + i\sigma/\omega\varepsilon_0. \tag{3}$$

Where $\sigma = \mathrm{Re}\,\sigma + \mathrm{Im}\,\sigma$, $\varepsilon_0$ is the permittivity of vacuum. Different degeneracy factor $g$ and effective background dielectric constant $\varepsilon_b$ represent different BDS for $\varepsilon_c = 3$. In this paper, the material of BDS is $Na_3Bi$ or $Cd_3As_2$ for $g = 4$, $\varepsilon_b = 12$ [25, 31]. Then we use the transfer matrix method (TMM) to calculate the reflection coefficient of the reflected light at the interface of the multilayer structure. When the incident light incident at $\theta$ along the z direction into the multilayer structure, the field coefficients of the electromagnetic wave at any position in that direction can be connected by the following matrix:

$$m_j(\Delta z, \omega) = \begin{pmatrix} \cos(k_{jz}\Delta z) & -\frac{i}{q_{jz}}\sin(k_{jz}\Delta z) \\ -iq_{jz}\sin(k_{jz}\Delta z) & \cos(k_{jz}\Delta z) \end{pmatrix}, \tag{4}$$

where $k_{jz} = k_0\sqrt{\varepsilon_j - \varepsilon_1 \sin^2\theta_i}$ is the component of wave vector propagating along z-axis in the medium, $q_{jz} = \sqrt{1 - \varepsilon_1 \sin^2\theta_i/\varepsilon_j}/\sqrt{\varepsilon_j}$ for p-polarization, $q_{jz} = \sqrt{1 - \varepsilon_1 \sin^2\theta_i / \varepsilon_j}$ for s-polarization, and $j = (1,2,3,...)$ express different material layers in the multilayer structure. So by connecting the electromagnetic wave field coefficient matrix for every different material layer, we can get the transmission matrix of the whole multilayer structure:

$$M(\omega) = \prod_{j=1}^{m} m_j(\Delta z, \omega) = \begin{pmatrix} M_{11}(\omega) & M_{12}(\omega) \\ M_{21}(\omega) & M_{22}(\omega) \end{pmatrix}. \tag{5}$$

According to the TMM, we can get the reflection coefficients of p-polarization and s-polarizationa

$$r_{p,s} = \frac{(M_{11} + M_{12}q_l)q_0 - (M_{21} + M_{22}q_l)}{(M_{11} + M_{12}q_l)q_0 + (M_{21} + M_{22}q_l)}, \tag{6}$$

where $M_{kt}(k,t=1,2)$ represents the element in matrix $M$. For p-polarization, $q_l = \cos\theta_l / \sqrt{\varepsilon_1}, q_0 = \sqrt{\varepsilon_1}/\cos\theta_i$, for s-polarization, $q_0 = \sqrt{\varepsilon_1}\cos\theta_i, q_l = \sqrt{\varepsilon_1}\cos\theta_l$. $\theta_l$ is the angle of emergence in the bottom of the structure we proposed. Further, we can obtain the reflectance $R_{p,s} = |r_{p,s}|^2$ for two kinds of polarization. The reflection coefficients can be written as follows based on Taylor series expansion of arbitrary spectrum components:

$$r_{p,s}(k_{ix}) = r_{p,s}(k_{ix}=0) + k_{ix}\left[\frac{\partial r_{p,s}(k_{ix})}{\partial k_{ix}}\right]_{k_{ix}=0} + \sum_{n=2}^{N} \frac{k_{ix}^N}{n!}\left[\frac{\partial^j R_{p,s}(k_{ix})}{\partial k_{ix}^j}\right]_{k_{ix}=0}. \tag{7}$$

Next, according to the angular spectrum theory we can obtain the spin transverse shift of the of the reflected light of the multilayer structure. we take the linearly polarized Gaussian beam as the incident light, and its angular spectrum expression is:

$$\tilde{E}_i(k_{ix}, k_{iy}) = \frac{\omega_0}{\sqrt{2\pi}} \exp\left[-\frac{(k_{ix}^2 + k_{iy}^2)\omega_0^2}{4}\right], \tag{8}$$

where $\omega_0$ is the beam waist, $k_{ix}$ and $k_{iy}$ are the wave vectors of incident light along the x and y directions, and the angular spectrum of reflected light beam can be expressed as:

$$\begin{bmatrix} \tilde{E}_r^H \\ \tilde{E}_r^V \end{bmatrix} = \begin{pmatrix} r_p & \dfrac{k_{ry}\cot\theta_i(r_p+r_s)}{k_0} \\ -\dfrac{k_{ry}\cot\theta_i(r_p+r_s)}{k_0} & r_s \end{pmatrix} \begin{bmatrix} \tilde{E}_i^H \\ \tilde{E}_i^V \end{bmatrix}, \qquad (9)$$

in the spin basis set, $\tilde{E}_i^H = (\tilde{E}_{i+} + \tilde{E}_{i-})/\sqrt{2}$, $\tilde{E}_i^V = i(\tilde{E}_{i-} - \tilde{E}_{i+})/\sqrt{2}$ represent the incident light in horizontal and vertical polarization states. $\tilde{E}_{i+}$, $\tilde{E}_{i-}$ represent the left- and right-handed circularly polarized components, respectively. $k_0 = \omega/c$ is the wave number in vacuum, $k_{ry}$ is the wave vector along the y direction of reflected light.

Furthermore, we can get the complex amplitude of reflected light for the reflected by Fourier transform:

$$E_{H,V}^{r\pm} = \int dk_{rx} dk_{ry} \tilde{E}_{H,V}^{r\pm} \exp\left[i\left(k_{rx}x_r + k_{ry}y_r + k_{rz}z_r\right)\right], \qquad (10)$$

at last, according to the centroid displacement formula, we can calculate the transverse shift of reflected light:

$$\delta_{H,V}^{\pm} = \frac{\iint y_r \left|E_{H,V}^{r\pm}\right|^2 dx_r dy_r}{\iint \left|E_{H,V}^{r\pm}\right|^2 dx_r dy_r}, \qquad (11)$$

in this paper, we only consider the zero-order Taylor series of reflection coefficient to calculate the transverse shift [28]:

$$\delta_H^\pm = \mp\left(1+\operatorname{Re}[r_s]/\operatorname{Re}[r_p]\right)\cot\theta_i/k_0, \qquad (12)$$

$$\delta_V^\pm = \mp\left(1+\operatorname{Re}[r_p]/\operatorname{Re}[r_s]\right)\cot\theta_i/k_0. \qquad (13)$$

## 3. Results and Discussions

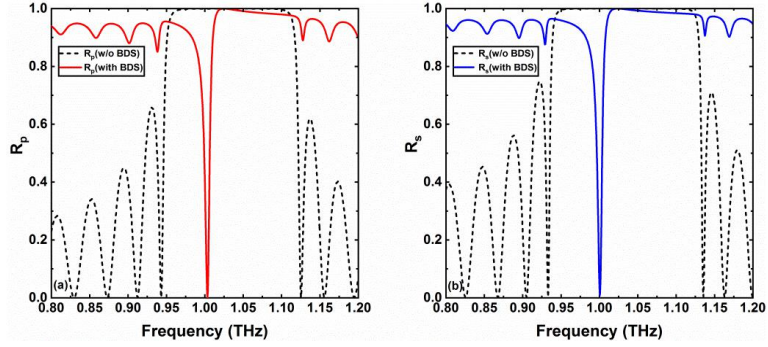

Fig. 2. The changes of the reflectance $R_p$ and $R_s$ near 1THz, the soild line indicate the structure with BDS, the dash line indicate the structure without BDS:(a) p-polarization state; (b) s-polarzation state.

We first discuss whether BDS can excite OTSs at terahertz range. It can be seen that without BDS in the structure, the multilayer structure cannot excite OTSs at both polarization conditions, so it only appears in the form of photonic band gap in the range of 0.95-1.13 THz. Since we ignore the dielectric loss, the reflectivity within the band gap is almost 100%. However, when BDS is added into the structure, a sharp reflection peak appears at 1THz due to the excitation of TPs between BDS and distributed Bragg reflector structure. The excitation condition of OTSs can be expressed by the following Eq: $r_{BDS}r_s\exp(2i\phi)=1$ [26]. $r_{BDS}=2/(1+\eta+\xi)-1$ is the reflection coefficient of incident light incident from the spacer layer to BDS, where $\eta=\varepsilon k_{2z}/\varepsilon_2 k_{1z}$, $\xi=\sigma k_{2z}/\varepsilon_0\varepsilon_2\omega$. $r_s$ is the reflection coefficient of incident light from the spacer layer to the photonic crystal. $\phi$ is the phase difference between the incident light propagating in the cavity between the two interfaces. We take p-polarization as an example to verify whether it satisfies the condition of excited OTSs by numerical calculation. When the incident light frequency is 1 THz and BDS is added into the structure, $r_s = 0.932 + 0.149i$, $r_{BDS} = 1.00 + 1.82 \times 10^{-6}i$, so $r_{BDS} r_s \exp(2i\phi) \approx 1$ is satisfied at this time, so the OTSs are excited. As a result, when the incident light frequency is 1THz, the reflectance for both polarization conditions produces sharp downward peaks. According to Eq. (12) and Eq. (13), We can see that the reflectance has an important effect on the size of the spin traverse, so we can boldly predict that when the frequency of incident wave is 1 THz, the addition of BDS layer can effectively excite OTSs, and the optical spin translocation is likely to appear a significant enhancement effect.

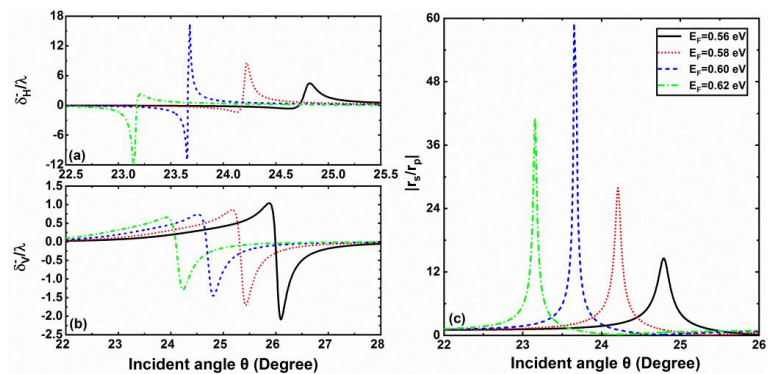

Fig. 3. The changes of transverse displacement (a) H-polarization state and (b) V -polarization state (c) value of $|r_s / r_p|$ for different Fermi energy, where $E_F = 0.56$ eV $0.58$ eV, $0.60$ eV, $0.62$ eV .

According to the results in Fig 2 above, when the frequency of incident light is 1 THz, the OTSs of the reflection interface of the multilayer structure can be effectively excited.Therefore, the incident light with a frequency of 1 THz is selected to specifically study the influence of Fermi energy of BDS on the transverse shift of the reflected light. As we predicted previously, both H-polarization and V-polarization have significant enhancement effect in a certain range of angles. It can also be seen from Fig. 3(a) and (b) that with the increase of Fermi energy, the peak of spin transversal shift generated by H-polarization and V-polarization is generated to a lower angle. It is worth mentioning that when $E_F = 0.60$ eV , the peak spin shift of H-polarization can reach about 16 $\lambda$ . According to Eq. (12), when value of $|r_s / r_p|$ increases, the transverse shift can be significantly enhanced, which is also well verified in Fig. 3(c), and the incidence angle of the peak value of $|r_s/r_p|$ is consistent with the peak value of transverse shift for H-polarization. It can be seen from Eq. (2) that the reason of Fermi energy can control the PSHE is that Fermi energy has a great influence on the imaginary part of the BDS dynamic conductivity. By adjusting the conductivity of BDS, we can effectively control the excitation intensity of TPs at the BDS interface to generate reflection peaks at specific incident angles, so as to enhance the PSHE. Therefore, we can adjust the Fermi energy of BDS by applying voltage according to this phenomenon, so as to achieve expediently control of PSHE at the terahertz range.

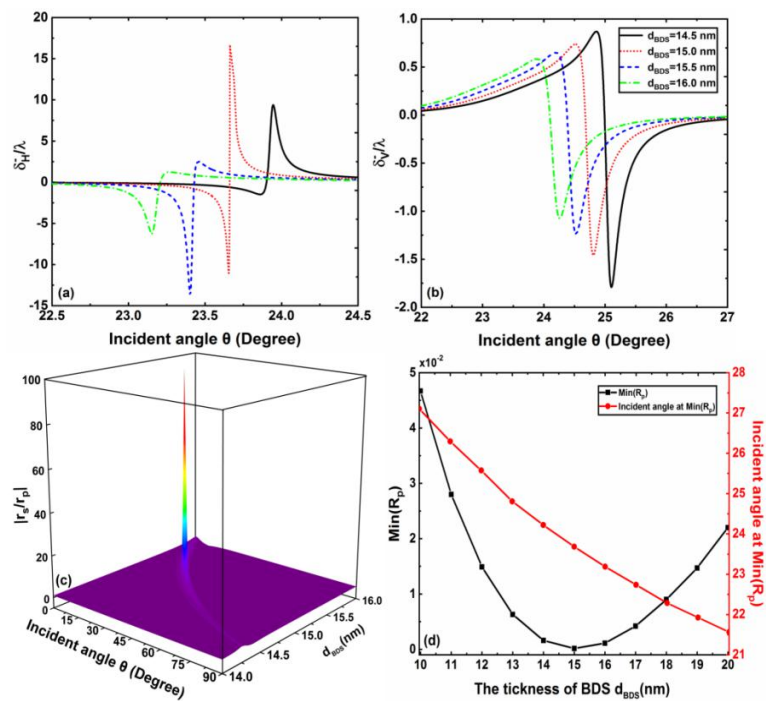

Fig. 4. The influence of BDS thickness on transverse shift (a) H-polarization state and (b) V-polarization .(c)The dependence of of the value of $|r_s/r_p|$ on different incident angles and tickness of BDS.(d) The influence of BDS thickness on $R_p$ minimum value and the incident angle when $R_p$ is minimum.

Next, we study the specific influence of structural parameters on spin transverse shift when the $E_F = 0.60$ eV, we plotted the change of spin transverse shift for H and

V-polarization with different thickness of BDS, as shown in Fig. 4(a) and (b). It can be seen that for H-polarization, when the thickness of BDS is 15 nm, the transverse shift can reach 17 times the working wavelength, for V-polarization, the transverse shift is smaller with the increase of $d_{BDS}$. It indicates that the intensity of Tamm plasmon excited by the multilayer structure is different for the two polarizations. Therefore, we can effectively control the PSHE by adjusting the thickness of BDS layer. In order to further explore the effect of BDS thickness on spin transverse shift, we plot a three-dimensional diagram of the effect of BDS thickness on reflection coefficient ratio $|r_s/r_p|$, as shown in Fig. 4 (c). It can be found that the reflection coefficient ratio $|r_s/r_p|$ increases significantly near the $d_{BDS}=15$ nm, so the PSHE for H-polarization is significantly enhanced. Moreover, it can be seen from Fig. 4(d) that $Min(R_p)$ achieves the minimum value when the BDS thickness is 15 nm, and with the increase of $d_{BDS}$, the incidence angle of generating the minimum value of $R_p$ presents a decreasing trend, which is consistent with the change of optical spin transverse shift with $d_{BDS}$ shown in Fig. 4(a) and (b). Therefore, it is necessary for us to study the law of the influence of the change of $d_{BDS}$ on the spin transverse shift, which will provide an important reference for us to design a reasonable optical spin device based on this structure.

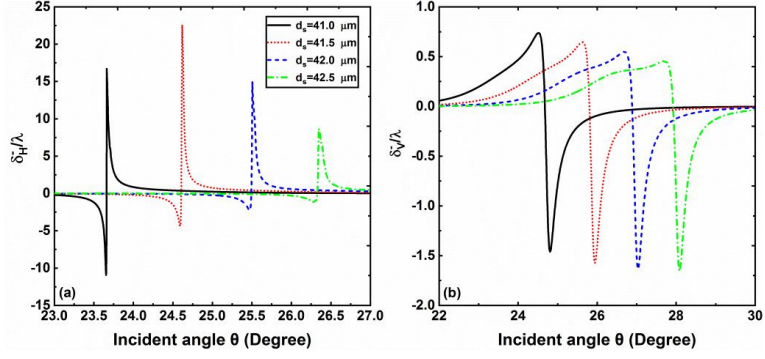

Fig. 5. The influence of space layer thickness on transverse shift, (a) H-polarization state and (b) V-polarization .

In order to further explore the enhancement of PSHE by adjusting other structural parameters, we find that the spacer thickness in the structure also has an important influence on spin transverse shift when $E_F = 0.60$ eV, $d_{BDS} = 15$ nm. It is not difficult to find from Fig. 5 (a) (b) that the spacer thickness $d_s$ can play an important role in controlling the PSHE. However, different from the influence of BDS thickness on the transverse shift, the angle of the optical spin transverse shift peak for the p and s-polarized waves will be greater with the increase of $d_s$. It can be seen from the previous conclusion that the reason for this phenomenon is closely related to the excitation of Tamm plasmon in the BDS-distributed Bragg reflector structure. Moreover, when $d_s = 41.5$ nm, the zero-order spin transverse shift for H-polarization can reach 22 times the working wavelength which provides another way for us to enhance and flexibly control the spin transverse shift when the incident light frequency is 1 THz.

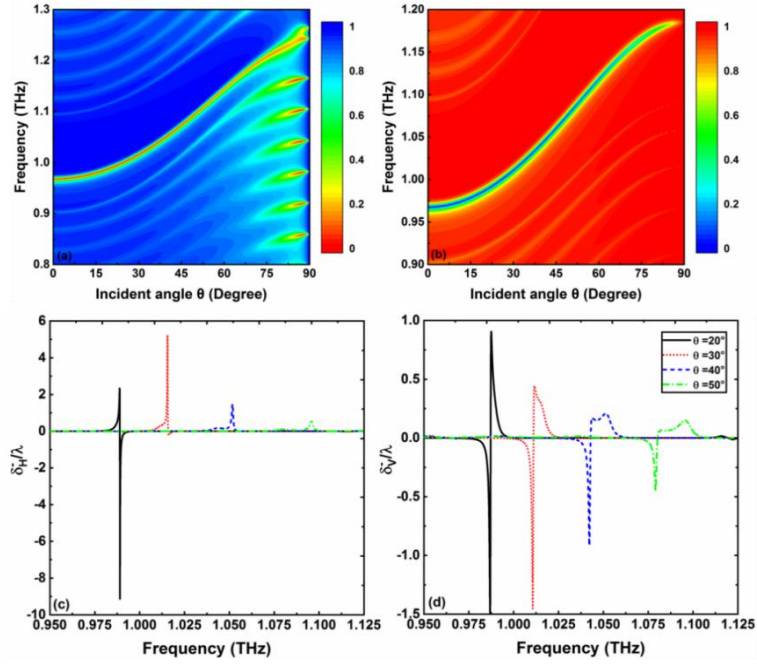

Fig. 6. (a),(b) Schematic diagram of the reflectance spectrum for p and s-polarization. The spin transverse shift at different frequencies when incident angle is 20°, 30°, 40°, 50°:(c) H-polarization state; (d) V-polarzation state.

From the Fig. 6(a) and (b), we can see that both p and s-polarization can generate continuous downward reflection peaks at different incident angles in the frequency range of 0.8-1.3 THz, which indicates that the two polarized waves in this frequency range can excite OTSs based on BDS-distributed Bragg reflector structure at different incident angles. In order to study the effect of this phenomenon on the optical spin transverse shift, as shown in Fig. 6(c)(d), we find that different incident angles do have a great influence on the spin transverse shift for two kinds of polarization. For the two polarization waves, the transverse shift decreases with the increase of incident angle. This shows that it is easier to obtain large transverse shift at lower incident angle and frequency near 1 THz. This is because the OTSs of the reflected light at the interface is excited more strongly at a lower incident angle and frequency, resulting in a larger reflection coefficient ratio. Therefore we believe that the incident angle also has an important influence on the controllable PSHE in different incident light frequency ranges, which provides a potential application for us to design non-contact controllable optical spin devices.

## 4. Conclusion

In conclusion, this paper theoretically studies the enhancement and control of the PSHE of the reflected light of Dirac semimetal distributed Bragg reflector structure in the terahertz range. The transverse shift of reflected light at the interface can be significantly enhanced as BDS can effectively excite OTSs in the terahertz range. The numerical simulation results show that with proper parameter optimization, the zero-order transverse shift can reach 2.5 times working wavelength for V- polarization and 22 times working wavelength for H-polarization. Moreover, the thickness of BDS and spacer layer is closely related to the enhancement of transverse shift. In addition, the Fermi energy of BDS plays an important role in the excitation of Tamm plasmon and the regulation of optical spin transverse shift, so we can control PSHE by adjusting the size of the external electric field. We also find that different incident angles can also control the PSHE in a certain frequency range. Therefore, we believe that the OTSs structure based on Dirac semimetal will provide potential application value for us to design enhanced and easily controllable optical spin devices at the terahertz range.

## Acknowledgments

This work was supported by the National Natural Science Foundation of China (Grant No. 11704119), the Hunan Provincial Natural Science Foundation of China (Grant No. 2018JJ3325), Scientific Research Fund of Hunan Provincial Education Department (Grant No. 21B0048) and National College Students' innovation and entrepreneurship training program (Grant No. 202110542014).